\documentclass[authoryear,preprint,review,12pt]{elsarticle}

\usepackage{amssymb}

\usepackage{lineno}

\journal{Icarus}

\begin{document}

\begin{frontmatter}


\title{Collisional heating and compaction of small bodies: \\ Constraints for their origin and evolution }


\author[label1]{Martin Jutzi}
\address[label1]{Physics Institute, University of Bern, NCCR PlanetS, Gesellschaftsstrasse 6, 3012 Bern, Switzerland}
\ead{martin.jutzi@space.unibe.ch}

\author[label2]{Patrick Michel}
\address[label2]{Universite C\^ote d’Azur, Observatoire de la C\^ote d’Azur, Centre National de la Recherche Scientifique, Laboratoire Lagrange, Nice, France. }




\begin{abstract}
The current properties of small bodies provide important clues to their origin and history. However, how much small bodies were processed by past collisions and to what extent they retain a record of processes that took place during the formation and early evolution of the Solar System is still poorly understood. 
Here we study the degree of collisional heating and compaction by analysing the large set of previous simulations of small body break-ups by \citet{Jutzi:2019jm}, which used porous targets of 50 - 400 km in diameter and investigated a large range of impact velocities, angles as well as energies. 
We find that the degree of impact processing is generally larger than found in previous studies which considered smaller objects \citep[e.g.][]{Jutzi:2017jbt,Schwartz:2018sm}. However, there is a clear dichotomy in terms of impact processing: the escaping material always experiences stronger heating than the material bound to the largest remnant. Assuming they originate from the same parent body, some of the observed differences between the recently visited asteroids Ryugu and Bennu may be explained by a different location of the material eventually forming these asteroids in the original parent body. Our results also provide constraints on the initial size of cometary nuclei.

\end{abstract}


\end{frontmatter}


\newpage

\section{Introduction}\label{s:intro}
The various objects composing the different small body populations, such as asteroids and comets, are the results of a billion-year-long collisional evolution of varying intensity. Their current properties provide important clues to their origin and history, and therefore represent a window to the early stages of the formation of the Solar System.
Most asteroids smaller than about 50 km in diameter are the result of a break-up of a larger parent body \citep{Bottke:2005bd}. They have been modified by subsequent collisions and their shapes, interior structure and spin state are determined to a large degree by the last major (global scale) impact event. 
Collisional processes - ranging from low-velocity mergers to catastrophic disruptions - have been proposed to also play an important role in the evolution of Kuiper Belt Objects and cometary nuclei \citep[e.g.][]{Durda:2000de,Morbidelli:2015vm}, although they are generally less frequent and less intensive than collisions between asteroids.  

Important questions related to collisions are: at which level were small bodies processed by past collisions and to what extent do they retain a record of processes that took place during the formation and early evolution of the Solar System? By analysing the degree of material processing (e.g. heating and compaction) during impacts, constraints for the origin and evolution of the small body populations as a whole as well as of individual objects - such as the recently visited asteroids Ryugu and Bennu or comet 67P/Churyumov–Gerasimenko - can be determined. Here, we perform such an analysis from a large data set of small body collisions produced in a previous study in which the focus was on the size and ejection speed distributions of resulting fragments \citep{Jutzi:2019jm}, and not on thermodynamical and physical state aspects. 

Previous studies of impact heating in small body collisions \citep[e.g.][]{Davison:2010dc,Davison:2013do,Ciesla:2013cd} have considered a limited range of conditions, mainly focusing on head-on impacts. The effect of impact obliquity was studied by \citet{Davison:2014dc}. However, these studies did not distinguish between heated material bound to the largest remnant and the heated material which is escaping the system. Recently, \citet{Wakita:2019wg} performed head-on impacts onto non-porous objects to specifically investigate the fate of hydrous materials. The simulations of small body break-ups by \citet{Jutzi:2019jm}, which are analysed here, used porous targets of 50 - 400 km in diameter and investigated a large range of impact velocities, angles and energies. We investigate the heating and compaction produced in these impacts and study how much processed material is retained (i.e., is bound to the largest remnant). 

In a recent study by \citet{Schwartz:2018sm}, it was found that not much heating or compaction occurs during the disruption of comets of 10 km in size, except for the small ejecta that do not contribute to the reaccumulation process producing the hierarchy of aggregates composing the main fragment distribution. Analysing the degree of material processing for bodies of 50 - 400 km in size allows us to understand how such a processing depends on scale and to apply it to small comet formation as well as to the formation of asteroid families whose parent bodies were in this size range, with implications for their physical properties. For instance, the NASA mission OSIRIS-REx and the JAXA mission Hayabusa2 both found that their asteroid targets, Bennu and Ryugu, have an hydration band, which is much deeper for Bennu than for Ryugu \citep{Hamilton:2019hs,Kitazato:2019km}. It is thought that they come from asteroid families whose parent body is in the 100 km size range \citep{Walsh:2013wd}. When such an asteroid family is created, does the material experience enough heating that any original hydrous material would be lost? Answering this question puts important constraints on the origin and evolution of small hydrated asteroids. In the context of comet formation, the analysis of  heating and compaction in break-ups of 50 - 400 km objects allows us to address the question whether or not comets were born big  \citep{Youdin:2005yg,Johansen:2007jo,Cuzzi:2010iv,Morbidelli:2009dd}), and what is their maximal initial size. Can small comets such as 67P originate from the disruption of such large bodies, or must they come from smaller ones that experience less heating as found in previous studies \citep{Jutzi:2017jb,Schwartz:2018sm}?

Here we concentrate on the processing of 1) material that remains bound to the largest remnant of a collisional disruption and 2) all the escaping material. In a future study, we will analyze the formation of aggregates and the fraction of heated and compacted material in these aggregates using a N-body code as in \citet{Schwartz:2018sm}. This will allow for a more complete understanding of how heated and compacted materials are distributed in the whole set of fragments (reaccumulated and escaping ones) produced by a large disruption. The possible modification of the interiors by early thermal evolution is investigated in a separate study (Golabek and Jutzi, in preparation).

In Sec. 2, we outline our modeling approach and in Sec. 3, we present the results of our analyses of the whole set of impact simulations covering a wide range of impact energies and angles. The implications of the results are discussed in Section 4. 

\section{Model approach}\label{s:methods}
The numerical scheme that was used to produce the data set of simulations that is analysed here is described in detail in paper I \citep{Jutzi:2019jm}. Here we provide a short summary of our collision modeling and present our approach to calculate the degree of heating and compaction in the largest collisional remnant and the escaping fragments.

\subsection{Collision modeling}\label{s:colmodel} 
To model an asteroid disruption and subsequent reaccumulations, \citet{Jutzi:2019jm} used an SPH / N-body approach as introduced by \citet{Michel:2001mb,Michel:2003mb}. The early phases of the collisions were simulated using an SPH impact code \citep{Benz:1994ij,Benz:1995hx,Nyffeler:2004tz,Jutzi:2008kp,Jutzi:2015gb} that includes self-gravity as well as material models including strength, friction and porosity. The N-body code pkdgrav was then used to compute the dynamical evolution of the system to late time.

\citet{Jutzi:2019jm} used porous parent bodies with a nominal diameter of $D$ = 100 km. To study the scale-dependence of the results, they additionally investigated $D$ = 50, 150 and 400 km targets. For the $D$ = 100 km objects, the same matrix of impact conditions as used in \citet{Durda:2007db}  and \citet{Benavidez:2012ez} was explored covering a wide range of impact speeds (from 3 to 7 km/s), impact angles (from 15$^\circ$ to 75$^\circ$ with 15$^\circ$ increments)\footnote{We note that for a random impact in 3 dimensions, the probability function $dP$ for the impact angle $\theta$ is given by $dP = sin(2\theta)d\theta$, giving a maximum probability of impact at $\theta$ = $45^\circ$. For a discussion of quasi 2-dimensional systems (e.g. flat disks), see \citet{Leleu:2018lj}} and impactor diameters, allowing a comparison with those two studies. In these studies, reaccumulating N-body particles were simply merged, as the main focus was on the fragment size and ejection speed distributions. The resulting shapes, the compaction of material and the distribution of heat were not investigated.

Here we focus on the degree of heating and compaction produced by the impact, which are all part of the outcome of the SPH simulations. In our analysis, we distinguish between the largest remnant and material not bound to this remnant. The  properties of the individual smaller fragments will be investigated in a future study.

\subsection{Determination of largest remnant and unbound fragments}

We analyse the simulation data obtained by \citet{Jutzi:2019jm} in a post-processing step in terms of heating and compaction of the largest remnant and the escaping fragments. The simulations performed in  \citet{Jutzi:2019jm} did not track the particle IDs in the N-body calculations, which would be required to follow the individual history of the fragments. Therefore, we compute the mass of the  largest remnant  $M_{lr}$ and the corresponding mass of the unbound fragments $M_f$ at the end of the impact phase using an iterative energy balance approach \citep{Benz:1999cj,Jutzi:2010bf}. This approach was demonstrated in previous studies to be consistent with the outcome of N-body simulations of the gravitational phase, as long as $M_{lr}$ is larger than a few 10\% \citep{Jutzi:2010bf} of the total mass. We apply this scheme to the simulations with specific impact energies $Q$ smaller than 2 times the catastrophic disruption threshold $Q^*_D$. For the small subset of simulations which have higher energies, we do not distinguish between largest remnant and smaller fragments. 

\subsection{Heating analysis}

The shock physics code calculations of the collisions were performed in \citet{Jutzi:2019jm} using the relatively simple Tillotson equation of state which does not allow for a direct computation of a thermodynamically consistent temperature.  However, we can still approximately compute the temperature increase $dT$ caused by an impact by analysing the increase in specific internal energy. As shown recently, this approach leads to temperatures comparable to the ones computed via the more sophisticated ANEOS equation of state, as long as vaporisation is negligible \citep{Emsenhuber:2018ej}. In Schwartz et al. (2018), it was assumed that the heat capacity is constant, while in reality it depends on temperature. To improve the realism, here we use a temperature dependent heat capacity to compute the temperature increase for rocky as well as cometary-like materials: 

\begin{equation}\label{eq:tc}
T = T_0 + \Delta u_i / c(T)
\end{equation}
 where $T_0$ is the initial temperature, $\Delta u_i$ the increase in specific energy and $c(T)$ the heat capacity. We then take into account the temperature dependence of the heat capacity by using the simple relation:  
 \begin{equation}\label{eq:ct}
c(T) = b + a T
\end{equation}
defined by the parameters $a$ and $b$. 
Using the equations (\ref{eq:tc}) and (\ref{eq:ct}) we obtain the following relation for the temperature as a function of specific internal energy:
\begin{equation} \label{eq:tu}
T = T_0 + \frac{\Delta u_i}{c(\Delta u_i)}
\end{equation}
with
\begin{equation}\label{eq:cu}
c (\Delta u_i) = \frac{1}{2} T_0 a+ \frac{1}{2} b+ \frac{1}{2} \sqrt{T_0^2 a^2 + 2 T_0 a b+b^2+4 \Delta u_i a}
\end{equation}

For rocky asteroids, we compute $c_{rock}=c(\Delta u_i)$ with $a$ = 2.5 and $b$ = 0, which provides a rough approximation to laboratory measurements of the heat capacity of forsterite at temperatures below 400 K \citep{Robie:1982rh}. These parameters, together with $T_0$ =  50 K, result in $c_{rock}$ = 1000 J/(kg K) at $T$ = 400 K (corresponding to $\Delta u_i$ = 3.5 $10^5$ J/kg). For $T >$ 400 K, we use a constant $c_{rock}$ = 1000 J$/$(kg K). 

For our analysis of cometary-like material we use a weight-averaged heat capacity \citep{Davidsson:2016ds}:
\begin{equation}\label{eq:cmix}
c_{mix}= f_{rock} c_{rock}+f_{ice} c_{ice}
\end{equation}
where $f_{ice}=1-f_{rock}$, $f_{rock}=m_{r,rock-ice}/(1+m_{r,rock-ice})$ and we assume a rock-to-ice mass ratio of $m_{r,rock-ice}$ = 2. The heat capacity of water ice $c_{ice}$ is computed using the parameters $a$ = 7.49, $b$ = 90 Ws/kg/K \citep{Klinger:1981}. In this analysis we also compute the mass fraction of sublimated water-ice for $T > $ 180 K as
\begin{equation}\label{eq:fsubl}
f = (\Delta u_i - u_{crit})/u_{vap}
\end{equation}
where  $u_{crit}$ = 188 kJ/kg is the specific energy at $T$ = 180 K (using $T_0$ = 50 K) and $u_{vap}$ = 3000 kJ/kg is the heat of vaporization \citep{Ahrens1984ao}.

\subsection{Compaction}

In the collision calculations by \citet{Jutzi:2019jm}, porosity was modeled using a sub-resolution approach based on the P-alpha model \citep{Jutzi:2008kp}. The material properties  (crush-curve) of the porous target used in this study were those that provided the best match to impact experiments on pumice targets \citep{Jutzi:2009ht}. The porosity model takes into account the enhanced dissipation of energy during compaction of porous materials.

In order to compute the degree of compaction caused by the collision, we consider the relative change of porosity, defined as 
\begin{equation}
\frac{p_0-p}{p_0} = \frac{\alpha_0-\alpha}{\alpha (\alpha_0-1)}
\end{equation}
where $p_0=1-1/\alpha_0$ is the initial porosity, $p=1-1/\alpha$ is the post-impact porosity, and $\alpha_0$ and $\alpha$ are the initial and post-impact distention, respectively. For full compaction, $(p_0-p)/p_0 = 1$ while in the case of no compaction, $(p_0-p)/p_0 = 0$.

\section{Results}\label{s:results}
\subsection{Impact heating of rocky material}
\begin{figure}
\begin{center}
\includegraphics[width=12cm]{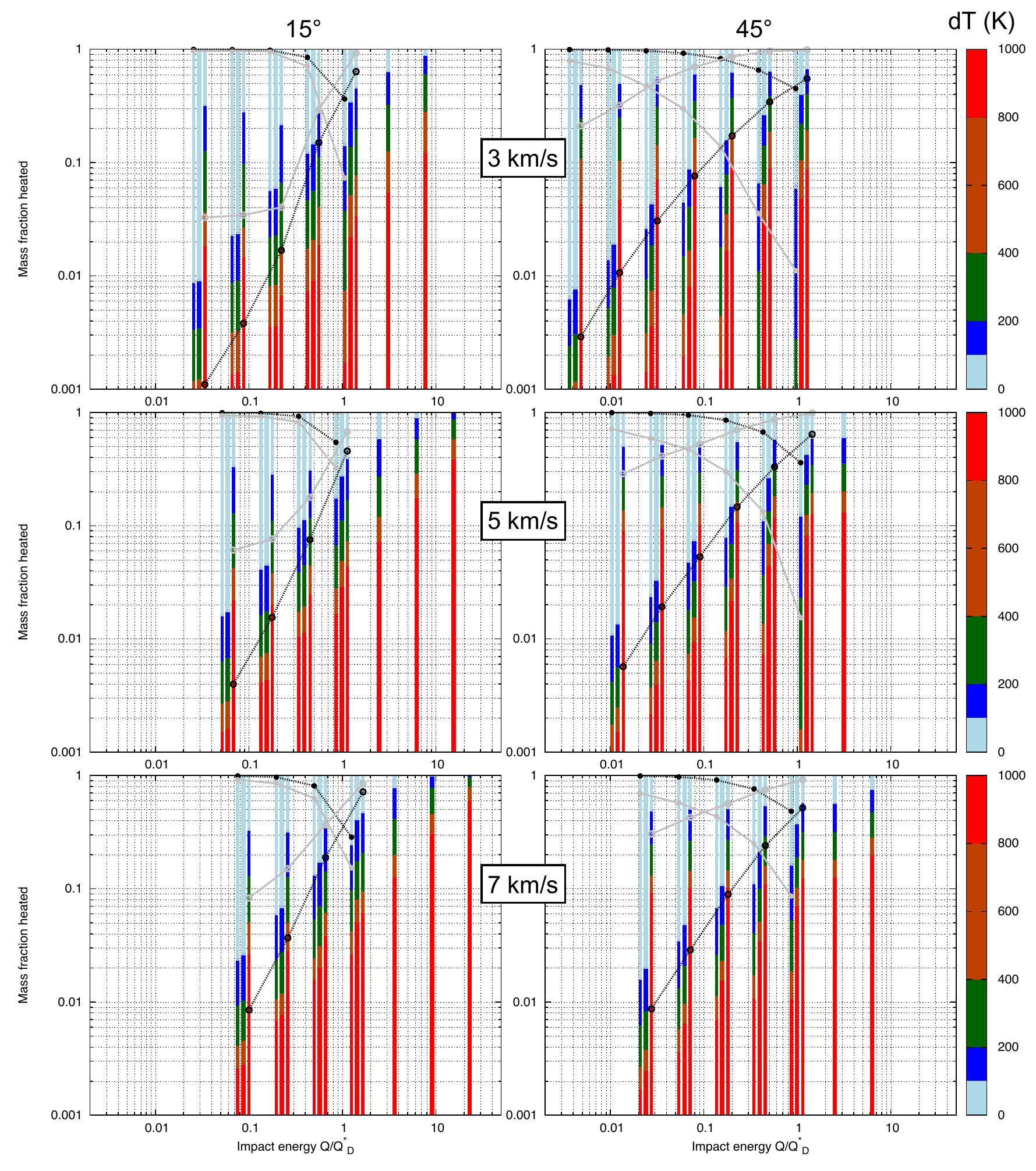}
\caption{Degree of heating in simulations using $D$ = 100 km targets, various impact angles (15$^\circ$ and 45$^\circ$) and velocities (3, 5, 7 km/s). In each case the mass fraction of material heated above a certain temperature for bound material (left column), all material (middle column) and escaping fragments (right column) is shown. The left and right columns are slightly shifted for a better visualisation. The x-axis corresponds to the specific impact energy $Q$ normalised by the catastrophic disruption threshold $Q^*_D$ (which is computed based on the data by \citet{Jutzi:2019jm} for each combination of impact angle and velocity). For $Q/Q^*_D > 2$, only the middle column is shown. The black line with filled symbols indicates the mass fraction of the largest remnant, while the black line with open symbols corresponds to the mass fraction of the escaping material. The gray lines show the corresponding fraction of thermal heat (filled symbols: contained in largest remnant; open symbols: contained in escaping material).}
\label{fig:tempdistr}
\end{center}
\end{figure}

The degree of impact heating of rocky material computed according to Eq.~(\ref{eq:tu}) is shown in Fig.~\ref{fig:tempdistr} for $D$ = 100 km targets as a function of the normalized specific impact energy for three different impact velocities and two impact angles ($15^\circ$ and $45^\circ$). For this computation we use the post-impact state of the largest remnant and the unbound material, 400 s after the impact, i.e. when the shock wave has long passed (see section \ref{sec:conv_tests}). The temperature increase is computed via Eq.~(\ref{eq:tu}) using  $c_{rock}$. Also displayed by the black and gray curves are the mass fraction of the largest remnant $M_{lr}/M_{tot}$, the mass fraction of the escaping material $M_{esc}/M_{tot}=1-M_{lr}/M_{tot}$ as well as the fraction of thermal energy contained in the largest remnant and the ejected material, respectively. 

We find that the fraction of thermal energy in the largest remnant is always (much) smaller than the mass fraction of the largest remnant, while for the escaping material it is the opposite. In other words, there is a clear dichotomy in terms of impact processing: the escaping material always experiences stronger heating than the material bound to the largest remnant. It is worth pointing out that this is not only the case for small ratios of $Q/Q*_D$ (cratering regime), but also in catastrophic or super-catastrophic collisions, where more than half of the initial mass is dispersed. As can be seen in Fig.~\ref{fig:tempdistr}, the differences are larger for more oblique impact angles. 

\begin{figure}
\begin{center}
\includegraphics[width=14cm]{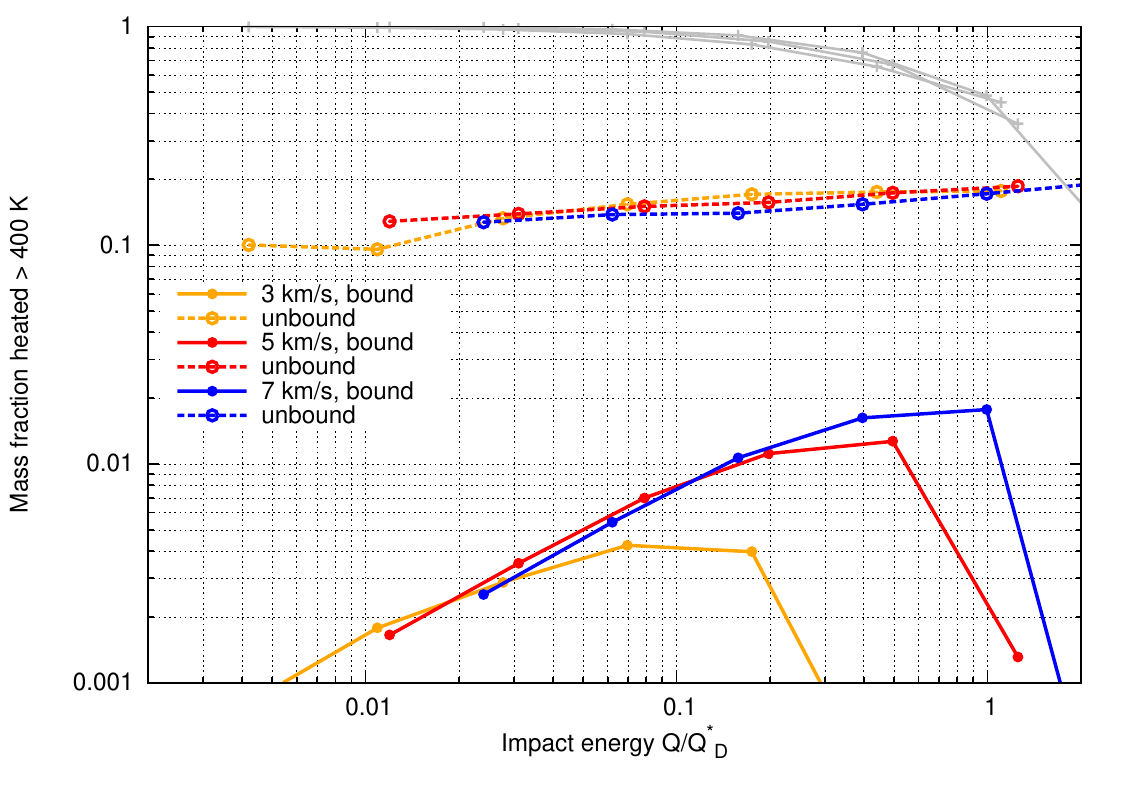}
\caption{Fraction of material heated by $dT>$ 400 K. Shown are the simulations with 45$^\circ$ impact angle and velocities of 3, 5 and 7 km/s. The gray lines show the relative sizes of the largest remnants.}
\label{fig:tempdistr400_45}
\end{center}
\end{figure}

\begin{figure}
\begin{center}
\includegraphics[width=14cm]{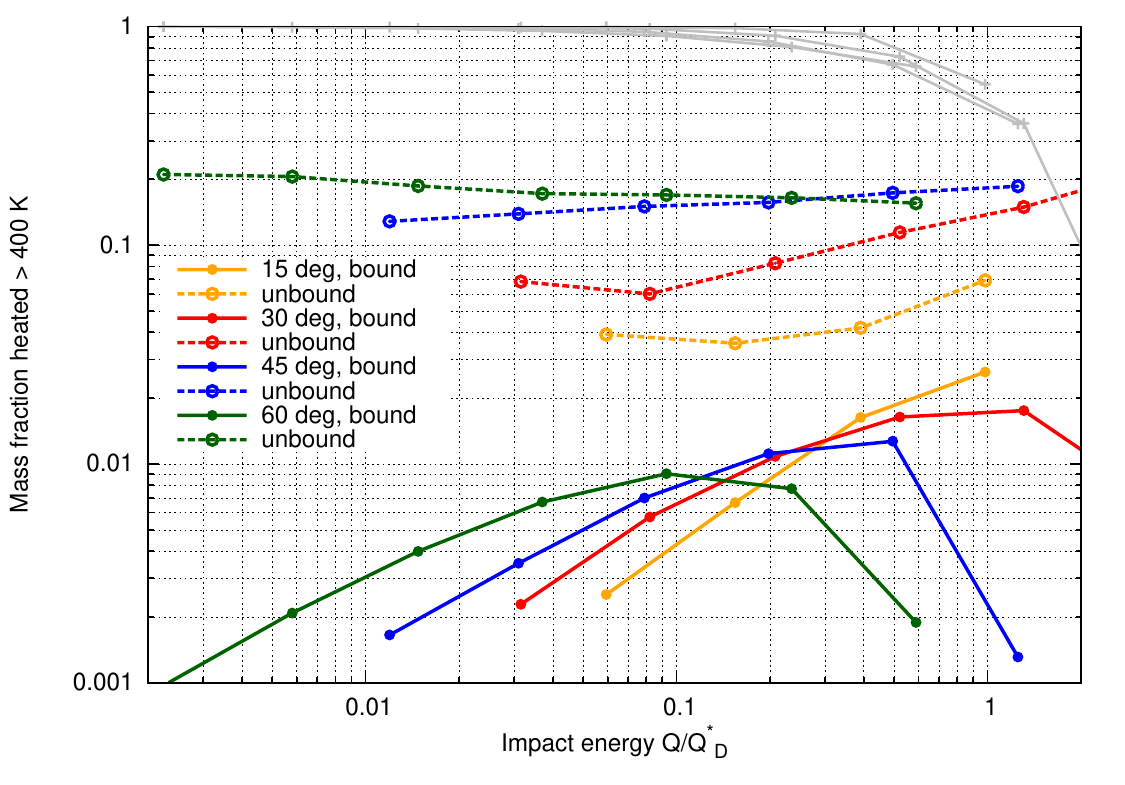}
\caption{Fraction of material heated by $dT>$ 400 K. Shown are the simulations with varying impact angles and a velocity 5 km/s. The gray lines show the relative sizes of the largest remnants.}
\label{fig:tempdistr400_5kms}
\end{center}
\end{figure}

Interestingly, for a given impact angle and velocity, the temperature increase experienced by the escaping material is roughly constant, as larger projectiles heat a larger region of the target, but at the same time eject a larger amount of target material. The heating of the bound material (largest remnant) increases with increasing impact energy, but is less than the heating of the escaping material. For the bound material, there is a maximum heating (which depends on impact angle and velocity) because the relative fraction of thermal heat contained in the largest remnant decreases faster with impact energy than its relative mass fraction (gray vs. black lines with filled symbols in Fig.~\ref{fig:tempdistr}). In other words, a larger fraction of heated material escapes at high impact energies. We also compute the mass fractions experiencing a specific temperature increase of  $dT>$ 400 K for a range of impact parameters (Fig.~\ref{fig:tempdistr400_45} and \ref{fig:tempdistr400_5kms}), which confirms these observations. 

\subsubsection{Convergence tests}\label{sec:conv_tests}
\begin{figure}
\begin{center}
\includegraphics[width=14cm]{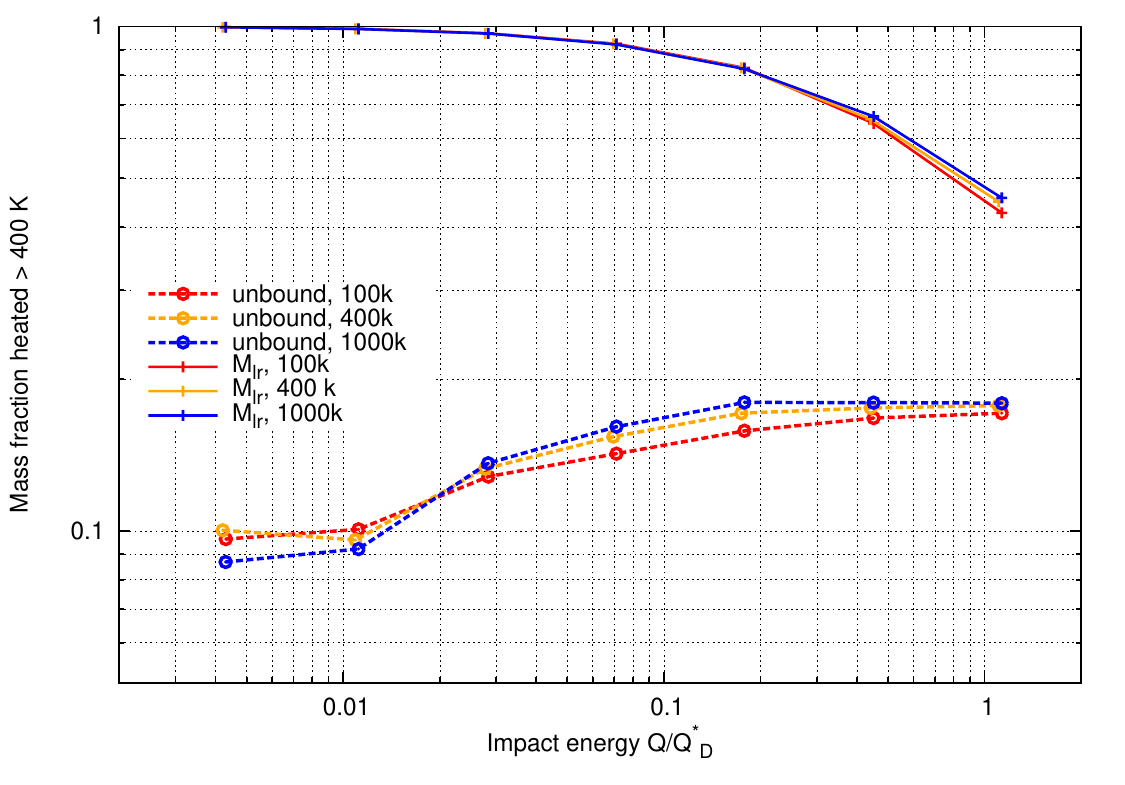}
\caption{Fraction of unbound material heated by $dT>$ 400 K and relative size of largest remnant ($M_{lr}$). Shown are the simulations with 45$^\circ$ impact angle and velocity of 3 km/s as a function of impact energy, using different numerical resolutions: 100 k, 400 k (nominal) and 1000 k SPH particles.}
\label{fig:tempdistr400_3_45_res}
\end{center}
\end{figure}

To quantify the dependence of our results on numerical parameters, the degree of impact heating and the mass of largest remnant are shown in Fig.~\ref{fig:tempdistr400_3_45_res} as a function of numerical resolution for the case  with 45$^\circ$ impact angle and velocity of 3 km/s using different projectile sizes and corresponding specific impact energies.

\begin{figure}
\begin{center}
\includegraphics[width=14cm]{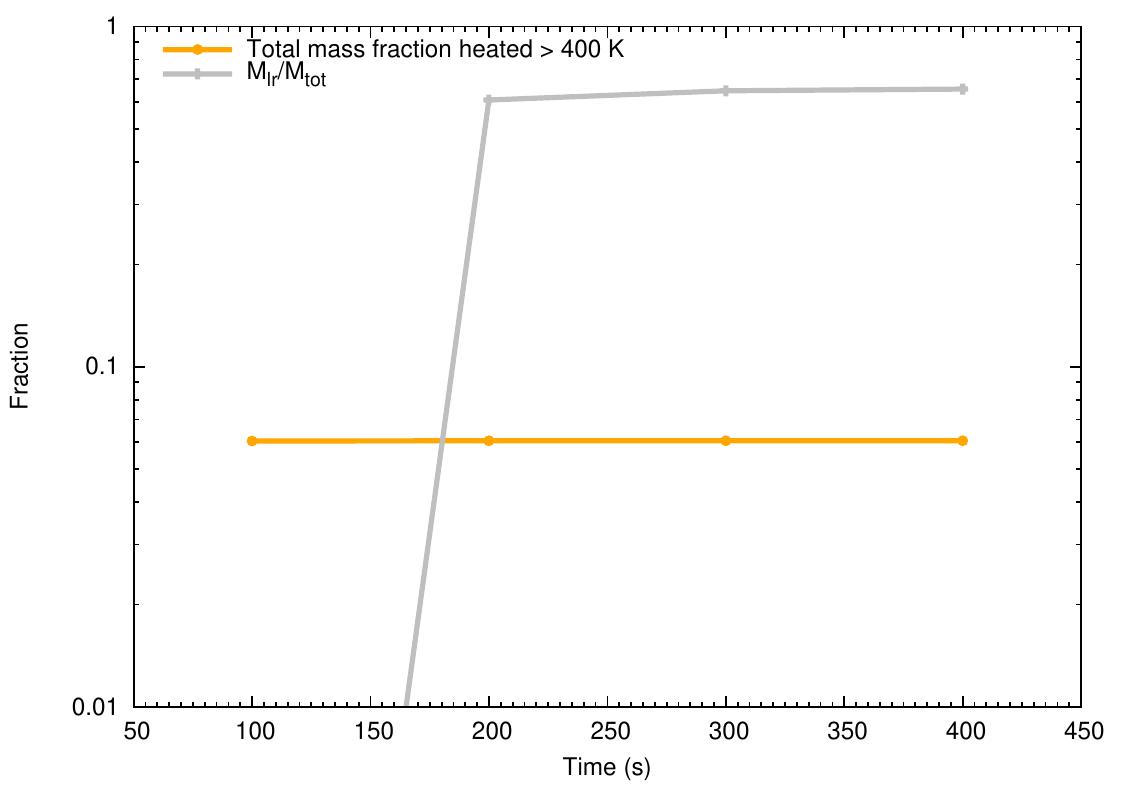}
\caption{Fraction of material heated by $dT>$ 400 K. Shown is the simulation with a 45$^\circ$ impact angle, a velocity of 3 km/s and a projectile radius of 13.4 km, using different times for the computation of the largest remnant and the degree of heating of the unbound material.}
\label{fig:tempdistr400_3_45_13_time}
\end{center}
\end{figure}

The dependence of the results on the time of the analysis is shown in 
Fig.~\ref{fig:tempdistr400_3_45_13_time}.

Overall, there is a reasonable convergence of the investigated quantities, especially given the potentially much larger effects of material properties, impact geometries, target structures and shapes, etc.

\subsection{Compaction}
\begin{figure}
\begin{center}
\includegraphics[width=14cm]{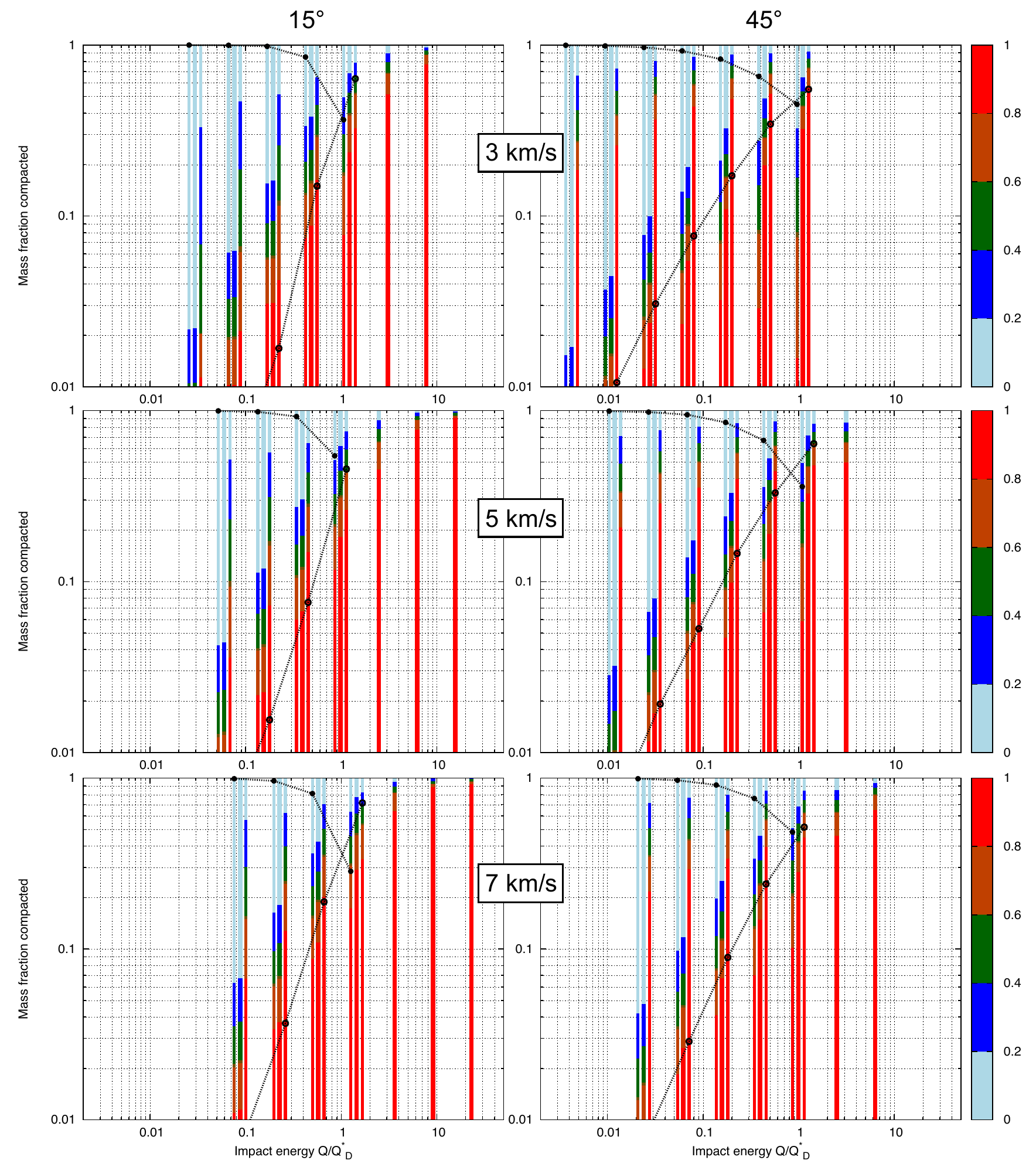}
\caption{Same as Figure \ref{fig:tempdistr}, but shown is the mass fraction of material compacted by a certain percentage (1: fully compacted; 0: no compaction). }
\label{fig:distdistr}
\end{center}
\end{figure}

\begin{figure}
\begin{center}
\includegraphics[width=14cm]{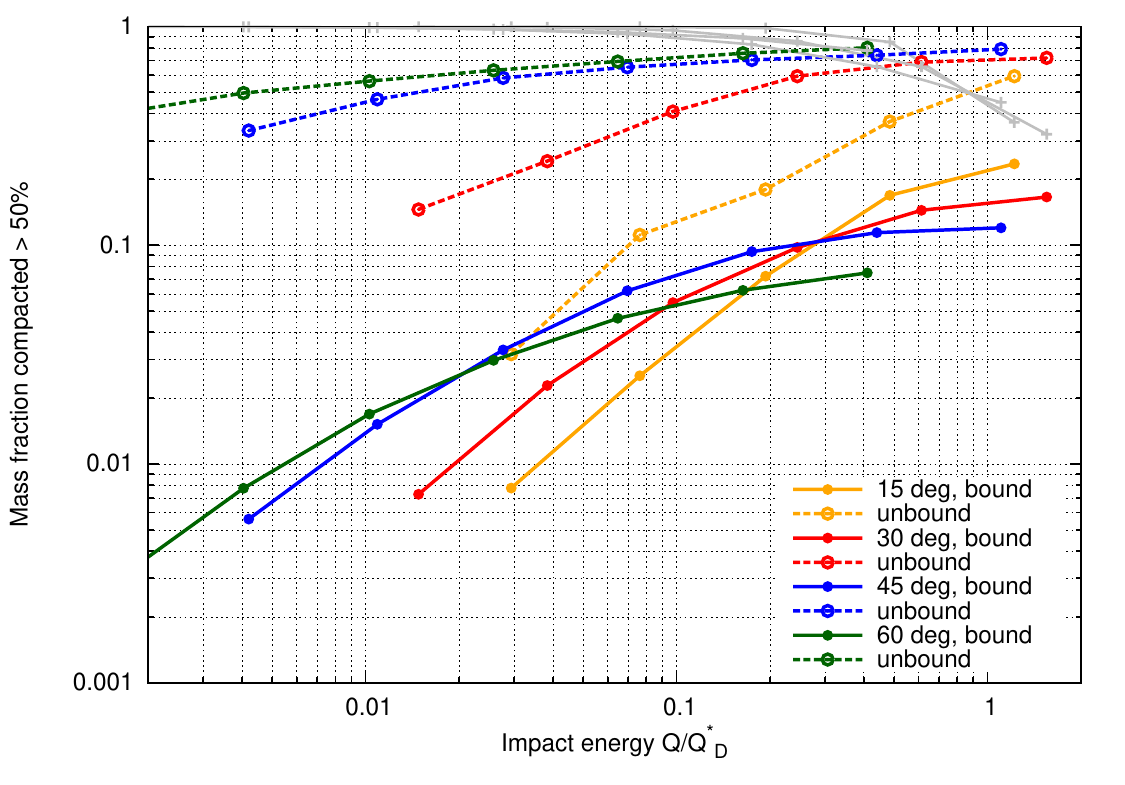}
\caption{Fraction of material compacted by more than 50\%. Shown are simulations with various impact angles and a velocity of 3 km/s.}
\label{fig:distdistr50_3kms}
\end{center}
\end{figure}

The degree of compaction (Fig.~\ref{fig:distdistr}) shows very similar characteristics as the degree of heating. The escaping material experiences a relatively high degree of compaction, independent of the impact energy, while the compaction of the bound material increases with increasing impact energy, but is much less significant. Figure \ref{fig:distdistr50_3kms} shows the mass fraction of material compacted by more than 50\% for specific cases. 
What is not taken into account in this analysis is the addition of macro-porosity by the reaccumulation process, which can again increase the overall porosity (as studied in \citet{Jutzi:2017jb}). 

\subsection{Impact heating of cometary-like material}
\begin{figure}
\begin{center}
\includegraphics[width=14cm]{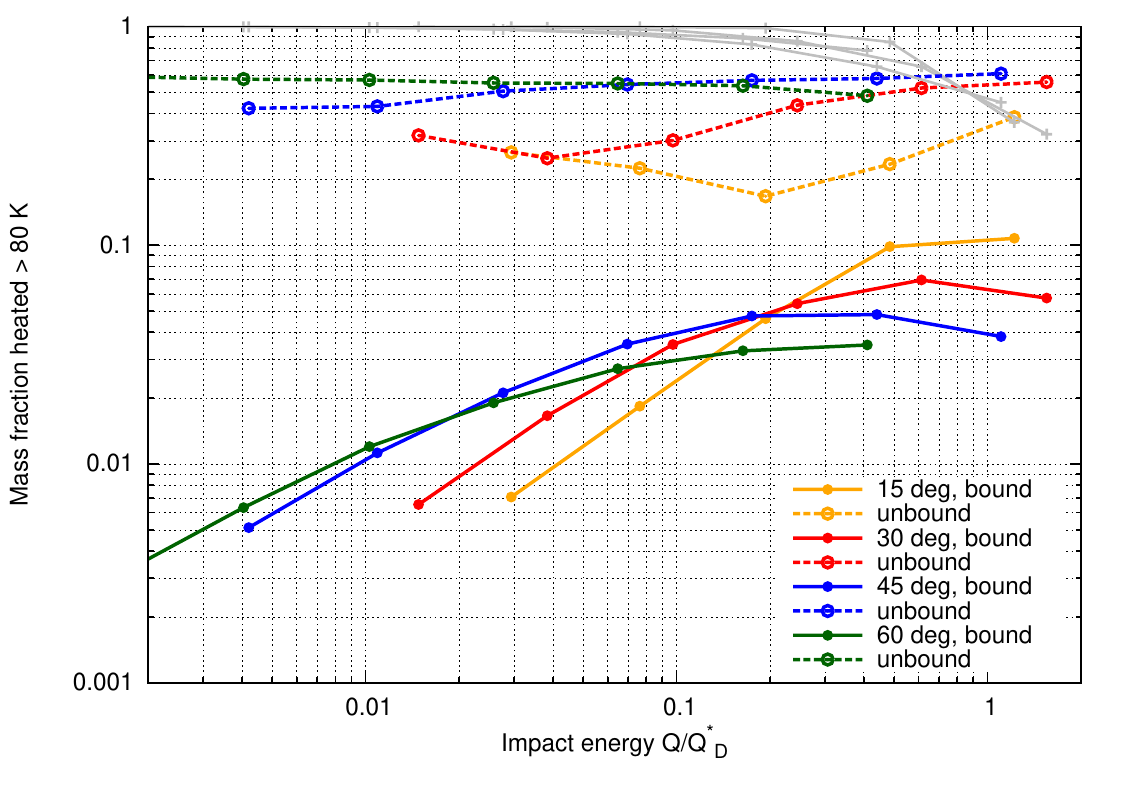}
\caption{Fraction of material heated by $dT>$ 80 K using cometary-like properties. Shown are simulations with various impact angles and a velocity of 3 km/s.}
\label{fig:tempdistr80_3kms}
\end{center}
\end{figure}

We also compute the temperature increase that a cometary-like material would experience due to impacts. To do this we use the heat capacity for a rock-ice mixture as defined in Eq.~(\ref{eq:cmix}). Figure \ref{fig:tempdistr80_3kms} shows the mass fractions of material heated by $dT>$ 80 K for the cases with 3 km/s impact velocity and varying impact angles.

\begin{figure}
\begin{center}
\includegraphics[width=14cm]{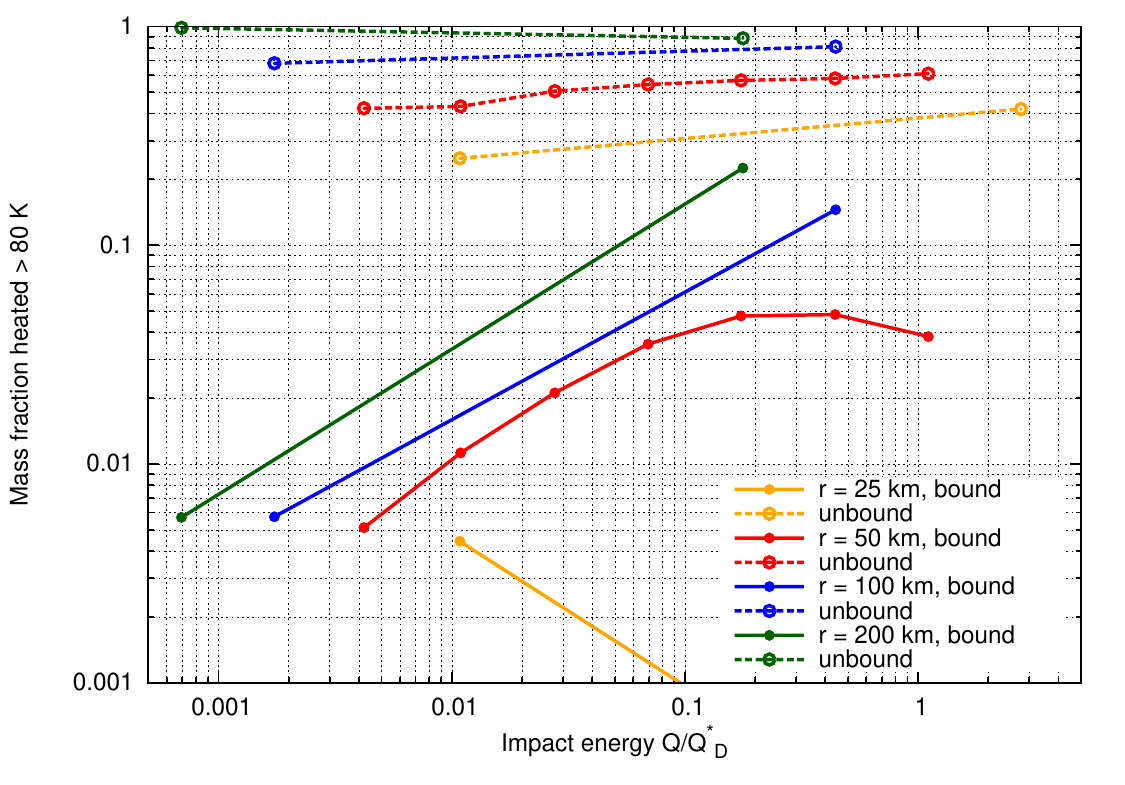}
\caption{Fraction of material heated by $dT>$ 80 K using cometary-like properties. Shown are the simulations with a 45$^\circ$ impact angle, a velocity of 3 km/s and varying target sizes.}
\label{fig:tempdistr80_45_3kms_varsize}
\end{center}
\end{figure}

\begin{figure}
\begin{center}
\includegraphics[width=14cm]{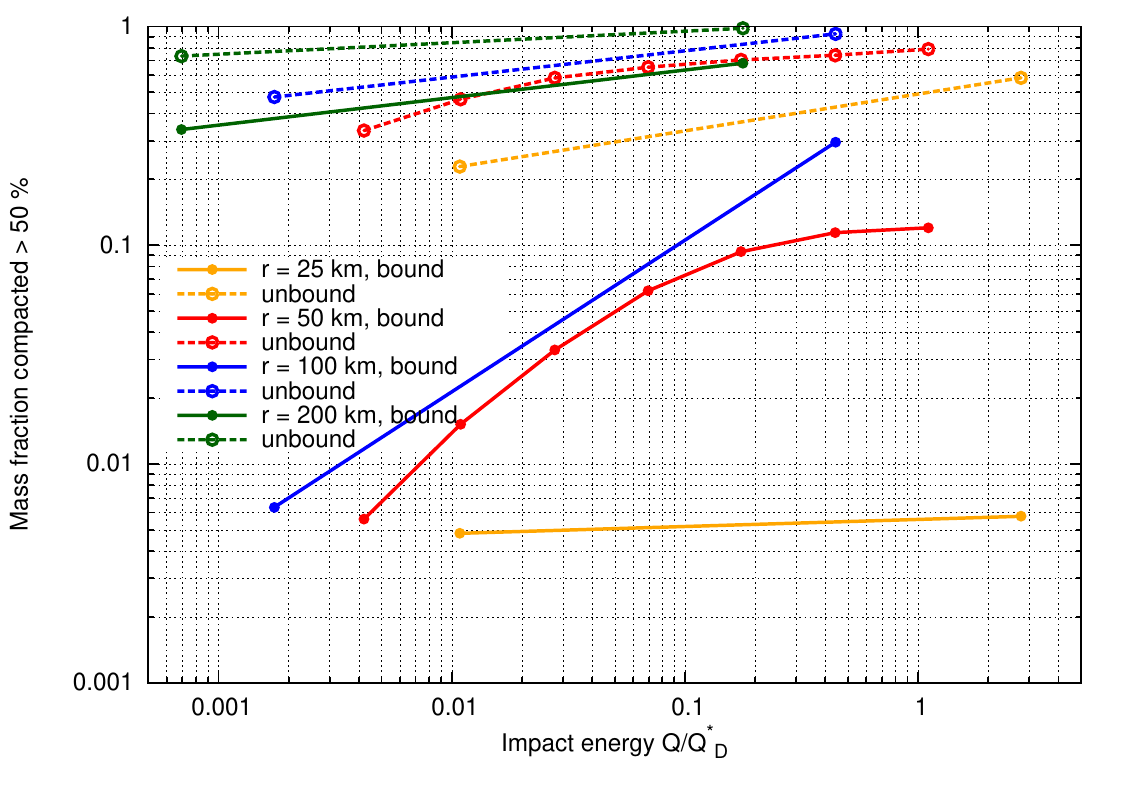}
\caption{Fraction of material compacted by more than 50\%. Shown are the simulations with 45$^\circ$ impact angle, a velocity of 3 km/s and varying target sizes.}
\label{fig:distdistr50_45_3kms_varsize}
\end{center}
\end{figure}

Most of the simulations in \citet{Jutzi:2019jm} were performed using $D$ = 100 km targets. However, a subset of impact parameters were studied using a range of targets sizes (50 - 400 km). These runs are used here to investigate the size dependence of the degree of heating and compaction. The results are summarised in Table \ref{table:tempdistr80_45_3kms_varsize} and displayed in Figs.~\ref{fig:tempdistr80_45_3kms_varsize} and \ref{fig:distdistr50_45_3kms_varsize}.

\begin{figure}
\begin{center}
\includegraphics[width=14cm]{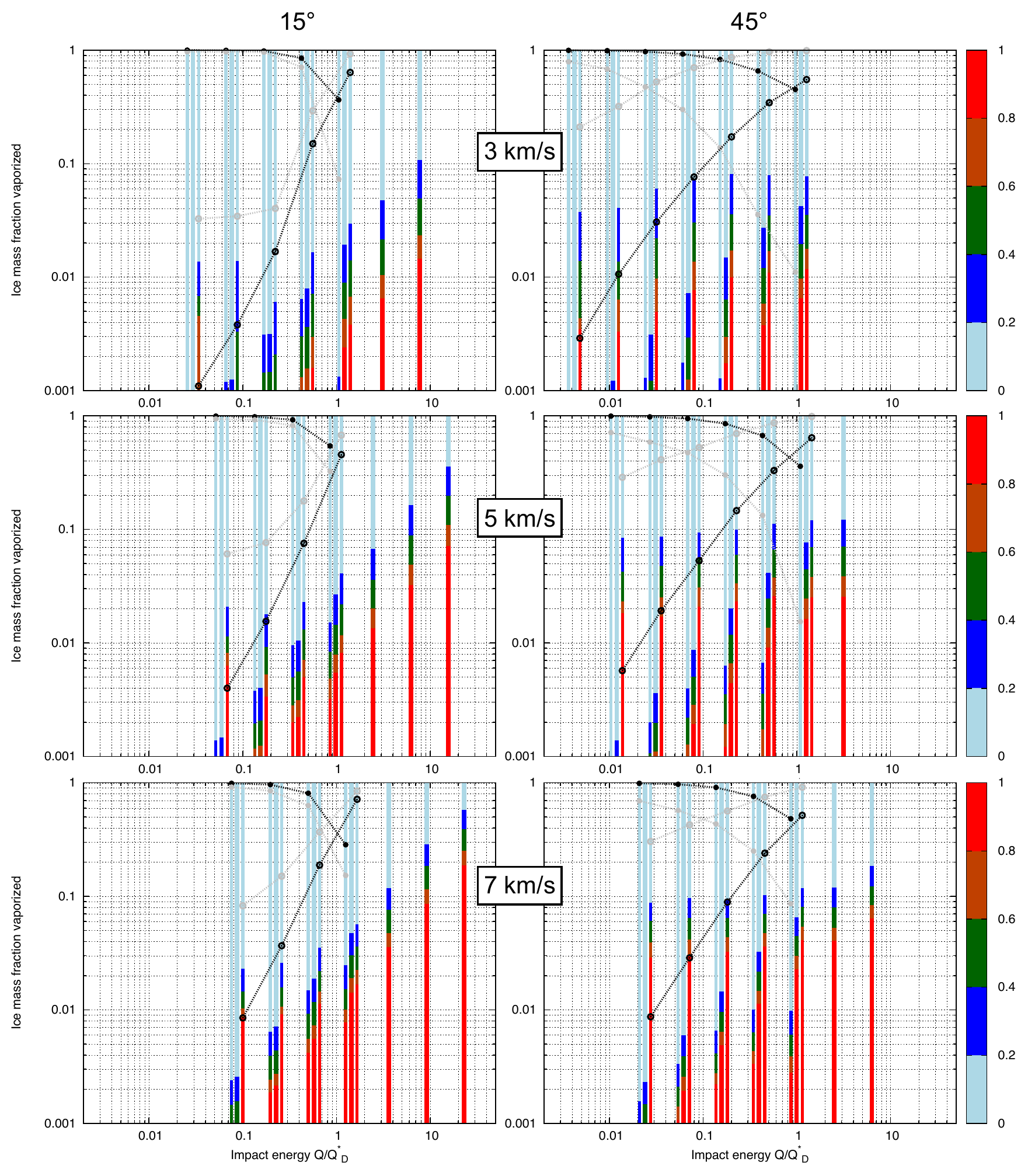}
\caption{Same as Figure \ref{fig:tempdistr}, but shown is the fraction of material (assuming cometary-like properties) with an ice vaporization fraction larger than a certain percentage (1: fully vaporized; 0: no vaporization).
}
\label{fig:vapdistr}
\end{center}
\end{figure}

The mass fraction of sublimated water-ice computed according to Eq.~(\ref{eq:fsubl}) is shown in Fig.~\ref{fig:vapdistr}. Except in the cases of the most energetic, super-catastrophic collisions, the impacts cause only little vaporization.

\section{Discussion and perspectives}
The results of our analysis of compaction and heating effects in collisional break-ups of 50 - 400 km sized bodies have important implications regarding the formation (initial size) and evolution of small solar system bodies. 

\subsection{Constraints for the formation of asteroids Ryugu and Bennu}

The asteroids Rygu and Bennu have been suggested to have originated from the Polana/Eulalia family forming events \citep{Walsh:2013wd,Sugita:2019sh}. With parent body sizes in the range of $D$ = 100 km, the masses of these subkm-sized objects are only $\sim 10^{-6}$ of the parent body mass. Current hybrid SPH--N-body simulations can therefore not properly resolve explicitly the formation of such small fragments (reaccumulated rubble piles), as they would only be represented by only 1-10 SPH particles in the original $D \sim$ 100 km parent body. However, our results (Fig.~\ref{fig:tempdistr400_45} and \ref{fig:tempdistr400_5kms}) suggest that the different degree of hydration observed on Ruygu and Bennu \citep{Sugita:2019sh,Lauretta:2019ld} does not necessarily mean that they come from different parent bodies, or that they experienced different degrees of dehydration after their formation, due for instance to their different perihelion histories. It could rather be due to the different degree of heating of the material that reaccumulated to form both of them in a single disruption event. In effect, depending on impact velocity and angle, up to 20\% of unbound material is heated $>$ 400 K (i.e., Ryugu could have been reaccumulated from this material), while the rest is heated less (i.e. this could be Bennu forming material). We note that given the small size of Ryugu and Bennu compared to the parent body, they are likely sourced from localized regions, which would be consistent with the uniform degree of hydration suggested by the spectral observations.

A more detailed analysis investigating the degree of heating in individual fragments and a discussion in the context of the formation of Ryugu and Bennu, including their shape, is presented in \citet{Michel:2020mb}.

\subsection{Contraints for the formation of comeatry nuclei}
Previous studies \citep[e.g.][]{Jutzi:2017jb,Jutzi:2017jbt,Schwartz:2018sm} found that at small $<$ 10 km scales, even catastrophic disruptions do not cause significant heating of the largest fragments. The collisions between $\sim$ 100 km sized objects analysed here take place at significantly larger scales. Because of the size-dependence of the catastrophic disruption energy, there is  significantly increased heating (as $Q^*_D$ strongly increases with increasing size). According to the scaling law $Q^*_D =a R^{3\mu}V^{2-3\mu}$ using $\mu=0.42$ \citep{Jutzi:2017jbt} and a constant velocity, $Q^*_{D, R = 50 km}/Q^*_{D,R = 3.5 km}$ $\sim$ 30, we can expect an overall increase of heating (compared to \citet{Schwartz:2018sm}) by the same factor. Indeed, our results (e.g., Fig.~\ref{fig:tempdistr80_3kms}) show significantly more heating than found by \citet{Schwartz:2018sm}  at smaller scales for comparable impact velocities (up to 1 km/s) and impact angles\footnote{We note that \citet{Schwartz:2018sm} used the same impact code, but slightly different material parameters, such as a lower density and crushing pressures.}. However, as discussed above, there is also a clear dichotomy in the degree of heating between the bound and unbound material.

For instance in the case of a 3 km/s impact with 45$^\circ$ angle and $Q/Q^*_D\sim 1.1$, we find that $\sim$ 4\% of the bound material experiences a temperature increase $dT >$ 80 K (i.e., 96\% is heated by less than 80 K). On the other hand, around 60\% of the unbound material experiences a temperature increase $dT >$ 80 K. For less disruptive impacts in the cratering regime with $Q/Q^*_D\sim 0.01$, we find that only $\sim$ 1\% of the bound material experiences a temperature increase $dT >$ 80 K, while still $\sim$ 40\% of unbound material has $dT >$ 80 K.

The analysis of the size dependence of these processes in collisional disruptions (Fig.~\ref{fig:tempdistr80_45_3kms_varsize} and \ref{fig:distdistr50_45_3kms_varsize} ; Table \ref{table:tempdistr80_45_3kms_varsize}) indicates that for bodies smaller than about $R$ = 25 km, the heating as well as the compaction of the bound material become negligible. This may suggest that the maximal initial size of cometary nuclei - assuming that comets were born big - must have been in this size range, in order for small comets resulting from the disruption of these  bodies to keep their primitive properties. 
	
However, the distribution of heat as a function of fragment size will require a more detailed analysis, following explicitly the evolution of the unbound material with a N-body code, as done by \citet{Schwartz:2018sm} and \citet{Michel:2020mb}, which will be performed more generally in a subsequent study. 

Another important aspect is the possible modification of the comet precursors' interiors by early thermal evolution (Golabek and Jutzi, in preparation). 

The complete understanding of small body heating needs to account for many aspects, and our study on impact heating and compaction, indicating in which case such processes are expected, provides direction for future investigations.

\section*{Acknowledgments}
The authors acknowledge funding support from the European Union's Horizon 2020 research and innovation programme under grant agreement No 870377 (project NEO-MAPP). M.J. acknowledges support from the Swiss National Centre of Competence in Research PlanetS. P.M. acknowledges funding support from the French space agency CNES as well as from Academies of Excellence: Complex systems and Space, environment, risk, and resilience, part of the IDEX JEDI of the Universit\'e C\^ote d’Azur. We also thank two anonymous reviewers for their constructive and valuable comments.



\bibliography{bibdata}



\newpage

\begin{table*}
\label{table:tempdistr80_45_3kms_varsize}      
\centering             
\begin{tabular}{l l l l l }        
\hline\hline            
 $R$ (km)	&	State	&	$Q/Q^*_D$	&	$f_{>80 K}$& $f_{>50\%}$	\\
 \hline        
25	&	bound	&	1.08E-02	&	4.43E-03	&	4.81E-03	\\
25	&	bound	&	2.76E+00	&	1.00E-04	&	5.76E-03	\\
25	&	unbound	&	1.08E-02	&	2.50E-01	&	2.29E-01	\\
25	&	unbound	&	2.76E+00	&	4.19E-01	&	5.83E-01	\\
50	&	bound	&	4.21E-03	&	5.11E-03	&	5.59E-03	\\
50	&	bound	&	1.09E-02	&	1.12E-02	&	1.52E-02	\\
50	&	bound	&	2.77E-02	&	2.12E-02	&	3.32E-02	\\
50	&	bound	&	6.95E-02	&	3.53E-02	&	6.19E-02	\\
50	&	bound	&	1.74E-01	&	4.75E-02	&	9.34E-02	\\
50	&	bound	&	4.41E-01	&	4.82E-02	&	1.14E-01	\\
50	&	bound	&	1.11E+00	&	3.83E-02	&	1.20E-01	\\
50	&	unbound	&	4.21E-03	&	4.23E-01	&	3.34E-01	\\
50	&	unbound	&	1.09E-02	&	4.32E-01	&	4.65E-01	\\
50	&	unbound	&	2.77E-02	&	5.07E-01	&	5.82E-01	\\
50	&	unbound	&	6.95E-02	&	5.43E-01	&	6.52E-01	\\
50	&	unbound	&	1.74E-01	&	5.68E-01	&	7.04E-01	\\
50	&	unbound	&	4.41E-01	&	5.79E-01	&	7.40E-01	\\
50	&	unbound	&	1.11E+00	&	6.10E-01	&	7.88E-01	\\
100	&	bound	&	1.74E-03	&	5.73E-03	&	6.34E-03	\\
100	&	bound	&	4.43E-01	&	1.46E-01	&	2.96E-01	\\
100	&	unbound	&	1.74E-03	&	6.79E-01	&	4.76E-01	\\
100	&	unbound	&	4.43E-01	&	8.09E-01	&	9.27E-01	\\
200	&	bound	&	6.95E-04	&	5.70E-03	&	3.38E-01	\\
200	&	bound	&	1.77E-01	&	2.25E-01	&	6.81E-01	\\
200	&	unbound	&	6.95E-04	&	9.88E-01	&	7.33E-01	\\
200	&	unbound	&	1.77E-01	&	8.84E-01	&	9.82E-01	\\

\hline                               
\end{tabular}
\caption{Fraction $f$ of material heated by $dT>$ 80 K (using cometary-like properties) and compacted by $> 50\%$ as a function of normalized impact energy $Q/Q^*_D$. The data corresponds to the simulations with a 45$^\circ$ impact angle, a velocity of 3 km/s and varying target radii $R$.}    
\end{table*}

\end{document}